\documentstyle[12pt,epsfig]{article}

\def\appendix{{\newpage\section*{Appendix}}\let\appendix\section%
        {\setcounter{section}{0}
        \gdef\thesection{\Alph{section}}}\section}

\newcommand{\non}{\nonumber \\}
\newcommand{\beq}{\begin{equation}}
\newcommand{\eeq}{\end{equation}}
\newcommand{\beqar}{\begin{eqnarray}}
\newcommand{\eeqar}{\end{eqnarray}}
\newcommand{\bea}{\begin{equation}}
\newcommand{\ee}{\end{equation}}
\begin{document}

\begin{titlepage}

\begin{flushright}
CERN-TH/2000-089\\
hep-th/0004011
\end{flushright}
\vfil\vfil

\begin{center}
{\Large {\bf  Graviton Scattering on D6 Branes with $B$ Fields}\\} 

\vspace*{15mm}
\vspace*{1mm}
Mohsen Alishahiha, Harald Ita and Yaron Oz 

\vspace*{1cm} 

{\it Theory Division, CERN \\
CH-1211, Geneva  23, Switzerland}\\

\vspace*{.5cm}
\end{center}

\begin{abstract}
We consider systems of D6 branes in the presence of a nonzero $B$ field of different
ranks.
We study the scattering of gravitons in the corresponding supergravity backgrounds.
We show that the nonzero $B$ field does not modify the form of the scattering potential.
The graviton scattering equation has two solutions one normalizable and one
non-normalizable. The normalizable solution does not lead to an absorption, however
the non-normalizable one does. We analyse the absorption of gravitons by the branes
and show that it is nonzero in the decoupling limit.
This result suggests that even in the presence of a $B$ field  
the D6 branes worldvolume theory does not decouple from the bulk gravity.
For comparison we analyse the form of the scattering potential and absorption for Dp branes
with $p <5$ and for NS5 branes.

\end{abstract}
\vskip 4cm

March 2000
\end{titlepage}

\newpage

\section{ Introduction}

The AdS/CFT correspondence (see \cite{OZ} for a review) relates 
field theories without gravity to supergravity (string) theories on certain curved
backgrounds.
The correspondence naturally arises when considering Dp branes in a limit
where the worldvolume field theory decouples from the bulk gravity \cite{Mal}. 
As discussed in \cite{HulDo}, when turning on a $B$ field on the Dp brane worldvolume the low energy effective 
worldvolume theory is deformed to a non-commutative Super Yang-Mills (NCSYM)
theory.
With  $N$  coinciding Dp branes in the presence of a nonzero $B$ field
the worldvolume theory is deformed to a $U(N)$ NCSYM \cite{SW}. 

Turning on a $B$ field on the Dp brane worldvolume can be viewed via the AdS/CFT correspondence
as a perturbation of the worldvolume field theory by a higher dimension operator.
The non-commutative effects are relevant in the UV and are negligible in the IR.
As in the cases with $B=0$, there exists a limit where the bulk gravity decouples
from the worldvolume non-commutative field theory \cite{CDS,SW}, and a correspondence
between string theory on curved
backgrounds with $B$ field and non-commutative field theories is expected.

With a vanishing $B$ field the worldvolume theory of $N$ D6 branes of Type IIA
string theory,
(in general Dp branes with $p >5$), does not decouple from the bulk.
There have been  indications that in the presence of a nonzero $B$ field
there is a limit where the worldvolume theory
of Dp branes with $p > 5$ may decouple from gravity.
In particular, for D6 branes with two non-commutative coordinates
there seemed to be for finite $N$ a UV description in terms of eleven dimensional supergravity
on a curved space, while for four or six  non-commutative coordinates
at finite $N$ a UV description in terms of ten dimensional supergravity
on a curved space \cite{AOS}.
On the other hand, the D6 branes system has a negative specific heat, which is
usually taken as a sign of non-decoupling of gravity.
Thus, the issue whether a  non-commutative
seven dimensional field theory on the worldvolume of D6 branes 
exists without gravity at all energy 
scales is still unsettled.

In this paper we will address this question by 
considering  systems of D6 branes in the presence of a nonzero $B$ field of different
ranks and analyse the scattering of gravitons in the corresponding supergravity backgrounds.
Generically, the sign of decoupling from the bulk is the vanishing of the absorption
cross section in the decoupling limit.
We will analyse the absorption of gravitons by the branes
and show that it is no-zero in the decoupling limit.
For comparison we analyse the form of the scattering potential and absorption for Dp branes
with $p <5$ (with and without a $B$ field), and for NS5 branes (with and without a RR 2-form
field), where worldvolume theories
 decoupled from the bulk exist.

The paper is organized as follows.
In the next section we review the decoupling limit with and without a $B$ field.
The analysis suggests a possibility of decoupling of D6 branes with a constant NS $B$ field on their worldvolume
from the bulk gravity. In the following  sections we will study this issue.
In section 3 we will
consider  systems of D6 branes with and without a $B$ field. 
We will analyse the scattering of gravitons in the corresponding supergravity backgrounds.
We will obtain the  graviton scattering equation and 
show that the nonzero $B$ field does not modify the form of the scattering potential.
We will then solve the graviton scattering equation. It  has two solutions one normalizable and one
non-normalizable. The normalizable solution does not lead to an absorption, however
the non-normalizable one does. 
We show that in the decoupling limit 
the absorption is not zero.
The result indicates the non decoupling of the bulk gravity from the brane modes.

The D6 branes supergravity background has a time-like singularity at the origin.
We will discuss this singularity and possible resolutions of it.
In section 4 we analyse the form of the scattering potential and absorption for Dp branes
with $p <5$ and for NS5 branes. Our interest in these cases is for comparison with the D6 branes cases.
The Dp branes supergravity backgrounds 
with $p > 3$ have a curvature singularity at the origin, and
when  $p < 3$ a dilaton blow-up.
In these cases the non-normalizable solution
of the graviton scattering equation leads to an absorption. However, we will see
clear
differences in the form of the scattering potential and the graviton absorption and their 
behaviour in the decoupling limit
compared to the 
D6 branes case. 

Other papers discussing various aspects of the supergravity backgrounds with $B$ field
in the context of dual descriptions of non-commutative gauge theories are
\cite{AKI,MALR,BR,HB,CH,KIM,DKT,AHNI}. Some absorption cross sections in the background of
 D3 branes with $B$ field have been computed in \cite{AK,MKL}. 


\section{The Decoupling Limit}

In the following we will review the decoupling limits for Dp branes with and
without a background of NS $B$ field \cite{Mal,OZ}. 
Consider a system of Dp branes extended along a $(p+1)$ dimensional plane in
$(9+1)$ dimensional space-time.
At low energies $E < 1/l_s$ only the massless string states are excited.
The bulk modes are the massless closed string states that include the graviton, and
the branes modes are the  massless open string states that include the gauge
fields.
Consider first the case with a vanishing $B$ field.
The leading terms in the interaction action between the brane modes and  
the bulk modes are obtained by covariantizing the brane action.
The quadratic term in the gauge field strength is 
\beq
S \sim \frac{1}{g_{YM}^2}
\int \sqrt{g}{\rm d}^{p+1}x g_{\alpha\beta}g_{\gamma\delta}F^{\alpha\gamma}
F^{\beta\delta} \ ,
\label{S}
\eeq
where $g_{YM}^2 = g_s l_s^{p-3}$ is the Yang-Mills gauge coupling. 
To be precise we have to add the 
dilaton field in (\ref{S}), however this does not affect the following discussion. 
We expand 
$g_{\alpha\beta} = \eta_{\alpha\beta} + \kappa_{10}h_{\alpha\beta}$ where
 $\kappa_{10} = l_{10}^4 = g_sl_s^4$, and $l_{10}$ is the ten-dimensional Planck
scale. In this notation $h$ is canonically normalized.

The action (\ref{S}) is now $S = S_{gauge} + S_{int}$ 
where 
\beqar
S_{gauge} &\sim& \frac{1}{g_sl_s^{p-3}}\int{\rm d}^{p+1}x 
F_{\mu\nu}F^{\mu\nu}  \ , \non
S_{int} &\sim&  \frac{1}{g_sl_s^{p-3}}\int{\rm d}^{p+1}x
( \kappa_{10} \eta_{\alpha\beta}h_{\gamma\delta}F^{\alpha\gamma}
F^{\beta\delta}  + O(h^2)) \ .
\label{sint1}
\eeqar
$S_{gauge}$ is the action of the worldvolume gauge theory and $S_{int}$ is the 
interaction action with the bulk gravity.

The decoupling limit of Dp branes 
from the bulk is the low energy limit.
In the decoupling limit we send $l_s \rightarrow 0$ and  keep the Yang-Mills coupling 
$g_{YM}^2 = g_sl_s^{p-3}$ fixed. When $p < 6$, the ten-dimensional
Planck length $l_{10} = g_sl_s^{1/4}$ vanishes  
implying that 
the interaction action (\ref{sint1}) vanishes. 
We should worry about   
the value of the eleven-dimensional Planck length
$l_{11} = g_s^{1/3}l_s$  as well since at
some energy scale we may need to pass to an eleven dimensional description of the system
where $l_{11}$ appears in the interaction action and not  $l_{10}$.
This happens for D2 branes at low energy and for D4 branes at high energy \cite{ITZ}.
When $p < 6$, 
the eleven-dimensional Planck length vanishes too in the above limit
implying that 
the interaction action vanishes in the eleven dimensional background. 
In these cases 
the field theories on the branes decouple from the 
bulk. 

When $p=6$ we keep the Yang-Mills coupling $g_{YM}^2 = g_sl_s^3$ fixed,
which means that we keep the eleven-dimensional Planck length
$l_{11}$ fixed. The phase diagram of the D6 branes system shows
that at high energy we have to use an eleven dimensional description \cite{ITZ}.
In view of the above discussion, 
the fact that $l_{11}$ is fixed means that interaction action does
not vanish and the bulk gravity does not decouple
from the field theory on the branes.
For NS branes the decoupling limit of from the bulk is the limit of weak coupling of the 
bulk physics $g_s \rightarrow 0$, while 
keeping $l_s$ fixed \cite{seiberg}. 

Consider now the case where we also have a background of constant NS $B$ field on the D branes
worldvolume.
In this case the decoupling limit is different.
In this set up, the end points 
of the open strings attached to the branes, $x_i$, are non commuting. 
Consider this system in the extreme condition
where $B_{i,i+1} \rightarrow \infty$ as  
$l_s\rightarrow 0$ such that $b_i \equiv l_s^2B_{i,i+1}$ is fixed.
Upon rescaling the coordinates $x_i \rightarrow \frac{b_i}{l_s^2} x_i$ and keeping the 
new coordinates fixed in the limit one gets 
\beq 
[x_i, x_{i+1}]=ib_i \ .
\label{nc}
\eeq
In the presence of the $B$ field, the massless states excitations of the open strings 
attached to the Dp branes 
give rise to a non-commutative worldvolume field theories, with
$b_i$ (\ref{nc}) being the deformation 
parameters.

Consider a $B$ field of rank $2m$. 
Now in the  decoupling limit we keep $g_{YM}^2 \sim g_sl_s^{p-3-2m}$ fixed.
Consider the D6 branes in the
presence of a $B$ field. 
In this case the rank of the $B$ field can be up to six,
$m=1,2,3$. 
When $m=1$ we need an eleven dimensional description 
in the UV \cite{AOS}.
Since in the decoupling limit we keep $g_sl_s=fixed$ as
$l_s\rightarrow 0$,
both the ten dimensional Planck length $l_{10}$ and 
the eleven dimensional Planck length $l_{11}$ vanish and it seems that
gravity decouples.
For $m=2,3$ the effective string coupling is small at all energy scales \cite{AOS}
and there are  
situations with no need for an eleven dimensional description at high energy.
Again,  both the ten dimensional Planck length $l_{10}$ and 
the eleven dimensional Planck length $l_{11}$ vanish
and it seems that gravity decouples.

We note however that the above argument can fail in the following way.
The theory on the branes is a non-commutative field theory which can be recast
as a commutative field theory with infinite number of terms in 
the gauge field strength and its derivatives \cite{SW}. 
With the inclusion of these terms the theory 
is non-local. We have been discussing 
the coupling to gravity as for a local theory, and analysis of the coupling
to gravity term by term suggests that the coupling is $l_{10}$ (or $l_{11}$).
This maybe misleading, and upon adding all the infinite number of terms it is possible
that the interaction of the complete non-local theory and gravity is not as simple.
As a comparison we can consider the issue of renormalizability of the non-local
non-commutative
field theory which seems surprising when viewed as a commutative
field theory with some higher dimension operators.

\section{Graviton Scattering on D6 Branes}

In this section we will study the scattering of gravitons on D6 branes with and without
a $B$ field. 
We will first compute the scattering potential and then analyse the scattering
in that potential.
We will start with the analysis of D6 branes with $B=0$ and continue with the cases 
of non-vanishing $B$ field with different ranks.

\subsection{D6 Branes with $B=0$}
We denote the ten dimensional coordinates by $x_0,x_1,...,x_9$.
Consider $N$ parallel D6 branes stretched in  $x_0,...,x_6$.
The  supergravity solution in the string frame takes the form  \cite{HS} 
\begin{eqnarray}
  ds^{2}&=&\sqrt{f(r)}(\frac{-dx_{0}^{2}+dx_{1}^{2}+dx_{2}^{2}+dx_{3}^{2}+dx_{4}^{2}+dx_{6}^{2}+dx_{5}^{2}}{f(r)}+
dr^{2} \nonumber\\
&+&r^{2}(d\vartheta^{2}+\sin^2\vartheta\, d\phi^{2})) \ ,\non
  e^{2\Phi}&=&f(r)^{-\frac{3}{2}} \ ,\non
  A_{\phi}&=&- R\,\cos\vartheta  \ ,\non
  f(r)&=&1+\frac{R}{r} \ ,
\label{D6}
  \end{eqnarray}
where $R \sim g_s l_s N$.
$\Phi$ is the dilaton and  $A_{\phi}$ is the RR 1-form dual to the  RR 7-form
that couples to the D6 brane worldvolume.

\begin{figure}[htb]
\begin{center}
\epsfxsize=4in\leavevmode\epsfbox{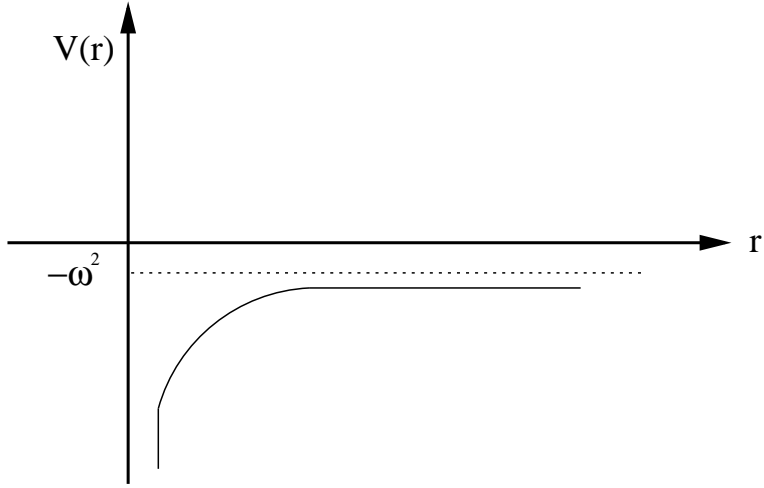}
\end{center}
\caption{The scattering potential $V(r)$
for gravitons polarized along the D6 brane.}
\label{D6fig}
\end{figure}

We will perturb the metric of the background (\ref{D6}) by 
\beq
g_{ab} = \bar{g}_{ab} + h_{ab},~~~~~~~~~a,b=0,...,9 \ ,
\label{per}
\eeq
where by $\bar{g}_{ab}$ we denote the background metric (\ref{D6}) and $h_{ab}$ is
the perturbation. The linearization of the Type IIA field equations is done
in appendix A. There are several polarizations $\varepsilon_{ab}$ 
of the graviton that have to be considered.
Also, we have to handle the ambiguities in the perturbation (\ref{per}) that arise
due to the diffeomorphism invariance by an appropriate gauge fixing.
We consider  s-wave gravitons with momenta along the brane 
\beq
h_{ab}=\varepsilon_{ab}h(r)e^{ik_{\mu}x^{\mu}},~~~~~~~ \mu=0,...,6 \ .
\eeq
The higher partial waves will simply encounter an additional centrifugal potential.
We choose the  gauge  
\beq
h_{a\mu}k^{\mu} = 0 \ ,
\label{gauge}
\eeq
that keeps transversal gravitons. 
We will choose the gravitons with polarization along the brane, i.e.
$\varepsilon_{ab}=0, a,b=7,8,9$. The other polarizations correspond to vectors and scalars
from the worldvolume theory point of view.
Let $k_{\mu}= w \delta_{0,\mu}$. We will consider other possibilities later.
This implies via (\ref{gauge}) that $\varepsilon_{a0}=0$. In addition there is a tracelessness
condition on the polarization tensor $\eta^{\mu\nu}\varepsilon_{\mu\nu} =0$ implied
by the linearized field equations. 
All together we see that there are 20 possible choices of polarizations of 
$\varepsilon_{\mu\nu}$. They can be realized by 15 
off-diagonal configurations such as 
\beq
\label{gravitymodes}
\varepsilon_{12}=\varepsilon_{21}=1,~~{\rm else}~~
 \varepsilon_{\mu\nu}=0 \ ,
\eeq
  plus 5 diagonal configurations such as
\beq
\varepsilon_{11}=-\varepsilon_{22}=1,~~{\rm else}~~
\varepsilon_{\mu\nu}=0 \ .
\eeq
Both types of polarizations yield the same equation for $h(r)$, which using the
linearized equations of appendix A and the background (\ref{D6}) reads
\begin{eqnarray}
  (\partial_{r}^{2}+a(r)\partial_{r}+b(r))h(r)=0 \ ,
\label{heq}
\end{eqnarray}
with the functions $a(r)$ and $b(r)$ given by 
\begin{eqnarray}
a(r)&=&\frac{2r+R}{r(r+R)} \ , \non
b(r)&=& \omega^{2}\left(1+\frac{R}{r}\right)-\frac{R^{2}}{4 r^{2}(r+R)^{2}}\ .
\label{ab}
\end{eqnarray}
We can write $h(r)=g(r)c(r)$ with  the function $c(r)$ given by 
\beq
c(r)=\frac{1}{\sqrt{r(r+R)}} \ .
\label{c}
\eeq
Now equation (\ref{heq}) can be recast as a schr\"{o}dinger-like equation for the function $g(r)$
and takes the form
\beq
(\partial_{r}^{2} - V(r)) g(r) = 0 \ ,
\label{sheq}
\eeq
with the potential $V(r)$ given by
\beq
  V(r)=-\omega^{2}(1+\frac{R}{r}) \ . 
\label{V}
\eeq
The potential (\ref{V}) is plotted in figure (\ref{D6fig}). It is an attractive Coulomb-like
potential. 
In order to get the scattering potential for the higher partial waves of the
graviton we simply have to add the
centrifugal piece, i.e. 
\beq
  V(r)=-\omega^{2}(1+\frac{R}{r}) + \frac{l(l+1)}{r^2} \ . 
\label{Vcent}
\eeq 
We will analyse the scattering in this potential later.

The analysis is rather general and can be easily modified for other perturbations
of the  D6 branes background.
For instance, if instead of a graviton we would have considered
 a minimally coupled scalar $\varphi(r)$, and decomposed it as 
 $\varphi(r)=g(r)c(r)$ with  the function $c(r)= 1/r$ 
then the equation for the function $g(r)$
would takes form (\ref{sheq}) 
with the potential (\ref{V}) for the s-wave mode and (\ref{Vcent}) for the higher partial
waves.
In general, the transversal graviton modes $h(r)$ are related to the minimally coupled scalar 
$\varphi(r)$ like
\begin{eqnarray}
  \label{eq:gravitonscalarrelation}
  h(r)&=&\frac{1}{\sqrt{f(r)}}\varphi(r) \ ,
\end{eqnarray}
with $f(r)$ being the harmonic function appearing in the metric, here given in (\ref{D6}).

\subsection{D6 Branes with rank two ($m=1$) $B$ field}

Consider $N$ parallel D6 branes stretched in  $x_0,...,x_6$, with a $B$ field
of rank two. We choose the $B$ field such that $B_{56} \neq 0$.
The  supergravity background
can be generated by starting with delocalized D5 branes oriented at an angle
in the $(x_5,x_6)$-plane and  
applying T-duality map on the $x_6$ coordinate. 
The background takes the form \cite{BGM}
\begin{eqnarray}
  ds^{2}&=&\sqrt{f(r)}(\frac{-dx_{0}^{2}+dx_{1}^{2}+dx_{2}^{2}+dx_{3}^{2}+dx_{4}^{2}}{f(r)}+\frac{dx_{6}^{2}+dx_{5}^{2}}{\sin^2\alpha_1+f(r)\cos^2\alpha_{1}}+dr^{2} \nonumber\\
&+& r^{2}(d\vartheta^{2}+\sin^2\vartheta\, d\phi^{2})), \non
  e^{2\Phi}&=&{\frac{1}{\sqrt{f(r)}( \sin^2\alpha_1+f(r)\cos^2 \alpha_1  ) }}, \non
  B_{65}&=&{\frac{(f(r)-1) \,\sin\alpha_1 \cos\alpha_1}
{\sin^2\alpha_1 +f(r)\, {{\cos^2 {{\alpha }_1}}}   }}, \non
  A_{\phi}&=&-R\,\cos \vartheta \,\cos {{\alpha }_1}, \non
  A_{56\phi}&=&-{\frac{R\,\cos \vartheta \,\sin {{\alpha }_1}}{\sin^2\alpha_1
+
f(r)\,\cos^2\alpha_1}},\non 
f(r)&=&1+\frac{R}{r} \ ,
\label{D61}
\end{eqnarray}
where $R \sim g_sl_s N \cos^{-1}\alpha_1 $.
In the limit $\alpha_1 \rightarrow 0$ the $B$ field vanishes and we recover the background
(\ref{D6}).  

There are now two cases to consider. 
Consider first a graviton polarized along the D6 brane but orthogonal
to the $B$ field.
The equation for the perturbation
$h(r)$ takes the form (\ref{heq}) with $a(r),b(r)$ given by (\ref{ab}).
Again, writing  $h(r)=g(r)c(r)$ with $c(r)$ given by (\ref{c}), 
the equation for $g(r)$ takes the form (\ref{sheq}) with the potential (\ref{V}).

Consider next a graviton polarized along the D6 brane parallel
to the $B$ field.
The equation for $h(r)$ takes the form
(\ref{heq}) 
with the functions $a(r)$ and $b(r)$ given by 
\begin{eqnarray}
a(r)&=&\frac{4r^2+6rR+R^2+R^2\cos{2\alpha_1}}{r(r+R)(2r+R+R\cos{2\alpha_1})},\nonumber\\ 
b(r)&=&\omega^{2}(1+\frac{R}{r})-\frac{R^{2}(-2r+R+(4r+R)\cos2\alpha_{1})}{4 r^{2}(r+R)^{2}(2r+R+R \cos2 \alpha_{1})} \ .
\label{abnew}
\end{eqnarray}
We can write $h(r)=g(r)c(r)$ where now $c(r)$ is given by
\beq  
c(r)=\frac{\sqrt{r + R}}{\sqrt{r}\left( 2\,r + R + R\,\cos 2 \alpha_1 \right)} \ .
\eeq
Now equation (\ref{heq}) can be recast as an equation for the function $g(r)$
and takes the form  (\ref{sheq}) with the potential (\ref{V}).

\subsection{D6 Branes with rank four ($m=2$) $B$ field}
Consider $N$ parallel D6 branes stretched in  $x_0,...,x_6$, with a $B$ field
of rank four. We choose the $B$ field such that $B_{34}$ and $B_{56}$ are
nonzero.
The  supergravity background
can be generated by starting with delocalized D4 branes oriented at angles
in the $(x_4,x_5,x_6)$-plane and  
applying twice the T-duality map, on the $x_5$ and $x_6$ coordinates. 
The background takes the form
\begin{eqnarray}
  ds^{2}&=&\sqrt{f(r)}(\frac{-dx_{0}^{2}+
dx_{1}^{2}+dx_{2}^{2}}{f(r)}+
\frac{dx_{3}^{2}+dx_{4}^{2}}{\sin^2\alpha_1+f(r)\,\cos^2\alpha_{1}} \nonumber\\
&+&
\frac{dx_{5}^{2}+dx_{6}^{2}}{\sin^2\alpha_2+f(r)\,\cos^2\alpha_{2}}+dr^{2}+ 
r^{2}(d\vartheta^{2}+\sin^2\vartheta\, d\phi^{2})),\nonumber\\
  e^{2\Phi}&=&{\frac{{\sqrt{f(r)}}}{( \sin^2\alpha_1+f(r)\, {{\cos^2 {{\alpha }_1}}}
   ) \,
     ( \sin^2\alpha_2 +f(r)\, {{\cos^2 {{\alpha }_2}}}  ) }},\nonumber\\
  B_{43}&=&{\frac{(f(r)-1 ) \,\sin {{\alpha }_1}\cos \alpha_1}
{\sin^2 \alpha_1 + f(r)\,{{\cos^2 {{\alpha }_1}}}   }}, \non
  B_{65}&=&{\frac{(f(r)-1 ) \,\sin{{\alpha }_2}\cos \alpha_2}
{\sin^2\alpha_2+f(r)\, {{\cos^2 {{\alpha }_2}}} }}, \non
  A_{\phi}&=&-R\,\cos \vartheta \,\cos {{\alpha }_1}\,\cos{{\alpha }_2}, \non
  A_{012}&=&-{\frac{\sin {{\alpha }_1}\,\sin {{\alpha }_2}}{f(r)}}, \non
  A_{56\phi}&=&-{\frac{R\,\cos \vartheta 
\cos {{\alpha }_1}\,\sin {{\alpha }_2}}{\sin^2\alpha_2 +f(r)\, {{\cos^2 {{\alpha }_2}}}  }}, 
\non
  A_{34\phi}&=&-{\frac{R\,\cos \vartheta \,\cos {{\alpha }_2}\,\sin {{\alpha }_1}}
{\sin^2\alpha_1 + f(r)\,{{\cos^2 {{\alpha }_1}}}   }}, \non
  f(r)&=&1+\frac{R}{r} \ ,
\label{D62}
\end{eqnarray}
where $R \sim g_sl_s N \cos^{-1} \alpha_1 \cos^{-1} \alpha_2 $.
In the limit $\alpha_1 \rightarrow 0$ the $B_{34}$ field 
vanishes and we recover the background
(\ref{D61}) with rank two $B$ field. 

Consider a graviton polarized along the D6 brane parallel
to the $B_{56}$ field.
The equation for $h(r)$ takes the form
(\ref{heq}) 
with the functions $a(r)$ and $b(r)$ given by 
\begin{eqnarray}
a(r)&=&\frac{4r^2+6rR+R^2+R^2\cos{2\alpha_2}}{r(r+R)(2r+R+R\cos{2\alpha_2})} , \non 
b(r)&=&\omega^2(1+\frac{R}{r})-\frac{R^2(-2r+R+(4r+R)
\cos{2\alpha_2})}{4 r^2(r+R)^2(2r+R+R \cos{2 \alpha_2})} \ .
\end{eqnarray}
As before, can write $h(r)=g(r)c(r)$ where now $c(r)$ is given by
\beq  
c(r)= \frac{\sqrt{r + R}}{\sqrt{r}\left( 2\,r + R + R\,\cos 2 \alpha_2 \right)} \ .
\eeq
and the equation for the function $g(r)$
takes the form  (\ref{sheq}) with the potential (\ref{V}).

\subsection{D6 Branes with rank six ($m=3$) $B$ field}
The highest rank for the $B$ field on the worldvolume of D6 branes
is six.
Consider now this case. All the previous cases are special limits of the current one.
Let again the $N$ parallel D6 branes be stretched in  $x_0,...,x_6$, with a $B$ field
of configuration of rank six, which we
choose such that $B_{12}$,$B_{34}$ and $B_{56}$ are
nonzero.
The  supergravity background
can be generated by staring with delocalized D3 branes oriented at angles
in the $(x_3,x_4,x_5,x_6)$-plane and  
applying three times the T-duality map, on the $x_4$, $x_5$ and $x_6$ coordinates. 
The background takes the form
\begin{eqnarray}
ds^{2}&=&\sqrt{f(r)}(\frac{-dx_{0}^{2}}{f(r)}+
\frac{dx_{1}^{2}+dx_{2}^{2}}{\sin^2\alpha_1+f(r)\,\cos^{2}
\alpha_{1}}+\frac{dx_{3}^{2}+dx_{4}^{2}}{\sin^2\alpha_2+f(r)\,\cos^{2}\alpha_{2}} \non
&+&\frac{dx_{5}^{2}+dx_{6}^{2}}{\sin^2\alpha_3+f(r)\,\cos^{2}\alpha_{3}}+dr^{2}+
r^{2}(d\vartheta^{2}+\sin^2\vartheta\, d\phi^{2})), \non
  e^{2\Phi}&=&\frac{f(r)^{\frac{3}{2}}}{\prod_{i=1}^3
( \sin^2\alpha_i + f(r)\,\cos^2{\alpha }_i) }, \non
  {B_{21}}&=&{\frac{(f(r)-1) \,\sin {{\alpha }_1}\,\cos 
\alpha_1}
    {\sin^2\alpha_1 + f(r)\,{{\cos^2 {{\alpha }_1}}}   }}, \non
  B_{43}&=&{\frac{(f(r)-1) \,\sin {{\alpha }_2}\,\cos \alpha_2}{\sin^2\alpha_2
+f(r)\, {{\cos^2 {{\alpha }_2}}}  }}, \non
  B_{65}&=&{\frac{(f(r)-1) \,\sin \alpha_3\cos \alpha_3}{\sin^2\alpha_3+f(r)\,
 {{\cos^2 {{\alpha }_3}}} }}, \non
 {A_0}&=&(\frac{1}{f(r)}-1)\sin 
{{\alpha }_1}\,\sin {{\alpha }_2}\,\sin {{\alpha }_3}, \non
 {A_{\varphi }}&=&-R\,\cos \vartheta \,\cos {{\alpha }_1}\,\cos {{\alpha }_2}\,\cos {{\alpha }_3}, \non
 {A_{012}}&=&{\frac{(f(r)-1) \,\sin {{\alpha }_2}\,\sin {{\alpha }_3}\, \cos \alpha_1}
    {\sin^2\alpha_1 + f(r)\,\cos^{2} {{\alpha }_1}   }}, \non
A_{12\,\varphi }&=&-
{\frac{R\,\cos \vartheta \,\cos \alpha_2\,\cos \alpha_3\,\sin \alpha_1 }
{ \sin^2\alpha_1 + f(r)\,\cos^2 \alpha_1   }}, \non
{A_{034}}&=&
{\frac{(f(r)-1) \,\sin \alpha_1\,\sin \alpha_3\, \cos \alpha_2}
    {\sin^2\alpha_2+ f(r)\,\cos^2 \alpha_2  }}, \non
{A_{34\,\varphi }}&=&-
{\frac{R\,\cos \vartheta \,\cos {{\alpha }_1}\,\cos {{\alpha }_3}\,\sin {{\alpha }_2} }
{ \sin^2\alpha_2 + f(r)\,\cos^{2} {{\alpha }_2}  }}, \non
{A_{056}}&=&{\frac{(f(r)-1) \,\sin {{\alpha }_1}\,\sin {{\alpha }_2}\, \cos \alpha_3}
    {\sin^2\alpha_3 +f(r)\, \cos^{2} {{\alpha }_3} }}, \non
{A_{56\,\varphi }}&=&
-{\frac{R\,\cos \vartheta \,\cos {{\alpha }_1}\,\cos {{\alpha }_2}\,\sin {{\alpha }_3}
          }{\sin^2\alpha_3 +f(r)\,\cos^{2} {{\alpha }_3} }}, \non
f(r)&=&1+\frac{R}{r} \ ,
\label{D63}
\end{eqnarray}
where $R \sim g_sl_s N \prod_{i=1}^3\cos^{-1} \alpha_i $.
In the limit $\alpha_1 \rightarrow 0$ the $B_{12}$ field 
vanishes and we recover the background
(\ref{D62}) with rank two $B$ field. 

As a more general example, 
consider now a graviton 
polarized parallel to the $B_{56}$ field  with momenta $k_1$ and $k_2$
along $B_{12}$ and $B_{34}$, respectively.
The equation for $h(r)$ takes the form
(\ref{heq}) 
with the functions $a(r)$ and $b(r)$ given by 
\begin{eqnarray}
a(r)&=&\frac{4r^2+6rR+R^2+R^2\cos{2\alpha_3}}{r(r+R)(2r+R+R\cos{2\alpha_3})},\non
b(r)&=&\left(\omega^{2}(1+\frac{R}{r})-k_{1}^2
(1+\frac{R\cos^2\alpha_1}
{ r})-k_{2}^2(1+\frac{R\cos^2\alpha_2}
{ r})\right)\non
&-&\frac{R^{2}(-2r+R+(4r+R)\cos{2\alpha_{3}})}{4 r^{2}(r+R)^{2}(2r+R+R \cos{2 \alpha_{3}})} \ .
\end{eqnarray}
As before, can write $h(r)=g(r)c(r)$ where now $c(r)$ is given by
\beq  
c(r)= \frac{\sqrt{r + R}}{\sqrt{r}\left( 2\,r + R + R\,\cos 2 \alpha_3 \right)} \ .
\eeq
The equation for the function $g(r)$
takes the form  (\ref{sheq}) with the potential
\begin{eqnarray}
V(r)=-\left(1+\frac{R}{r} \right) \,\omega^2+k_{1}^2
(1+\frac{R\cos^2\alpha_1}{ r})+k_{2}^2(1+\frac{R\cos^2\alpha_2}{ r})
\ ,
\label{kV}
\end{eqnarray}
with $\omega^2 -k_1^2 -k_2^2 \geq 0$.
We see that the momenta $k_1,k_2$ do not change the structure of the potential.
In particular, we can recast (\ref{kV}) in the form (\ref{V}) by 
redefining $\omega$ and $R$.

\subsection{Graviton Scattering}
Let us summarize the information gained in the above analysis. We considered 
a graviton polarized along the D6 brane worldvolume with
$k_{\mu}= w \delta_{0,\mu}$. We saw that if
we
decompose the graviton $h(r) = g(r)c(r)$ then with an appropriate choice of
$c(r)$ the equation for $g(r)$ takes the form (\ref{sheq}) with the potential
(\ref{V}) for the s-wave and 
(\ref{Vcent}) for the higher partial waves.
The only difference between having a non-vanishing $B$ field or not is reflected
in the form of the function $c(r)$. 
The differential equation (\ref{sheq}) with the
potential (\ref{Vcent}) can be solved exactly. It has two solutions
\beqar
g_1(r) &=& (\omega r)^{l+1}e^{i\omega r}\, _1F_1(l+1-{i\omega R \over 2},2+2l;-2i\omega r), \non
g_2(r) &=& (\omega r)^{l+1} e^{i\omega r} U(l+1-{i\omega R \over 2},2+2l;-2i\omega r) \ ,
\eeqar
where $_1F_1$ and $U$ are the Kummer confluent hypergeometric functions.
Consider the s-wave.
The asymptotic behaviour of the functions $g_1(r)$ and $g_2(r)$
as $r\rightarrow \infty$ and $r\rightarrow 0$ is
\beqar
g_1(r\rightarrow \infty) &\sim &  
e^{-i \pi(1-{i\omega R \over 2})} (-2i\omega r)^{{i \omega R \over 2}}
\frac{e^{i\omega r} }{\Gamma[1+{i\omega R \over 2}]} \non
&+& 
(-2i\omega r)^{-{i\omega R \over 2}}
\frac{e^{-i\omega r}}{\Gamma[1-{i\omega R \over 2}]}, \non
g_2(r\rightarrow \infty) &\sim &   (-2i\omega r)^{{i\omega R \over 2}} e^{i\omega r} \ ,
\label{exp1}
\eeqar
and 
\beqar
g_1(r\rightarrow 0) &\sim&   \omega r e^{i\omega r}, \non
g_2(r\rightarrow 0) &\sim&   \frac{-2i\omega r e^{i\omega r}}{\Gamma[-{i\omega R \over 2}]}\left[-\frac{1}{R\omega^{2}r}+\log{(2\omega r)}+\alpha_{1}\right] 
\ .
\label{exp2}
\eeqar
Where  $\alpha_{1}=\psi(1-\frac{iR\omega}{2})-i\frac{\pi}{2}-1+2e_{\gamma}$ with $\psi(x)=\partial_{x}\log{(\Gamma[x])}$ and $e_{\gamma}$ denotes the Euler-Gamma constant.

Both $g_1(r)$ and  $g_2(r)$ are regular solutions.
However, we should recall that the graviton function is  $h(r) = g(r)c(r)$.
As $r \rightarrow 0$ the function $c(r) \sim 1/\sqrt{r}$ independently of the rank 
of the $B$ field. Thus, while $h_1(r) = g_1(r)c(r)$ is regular at the 
origin, $h_2(r) = g_2(r)c(r)$
diverges there.   
Recall
that $h_{ab}=\varepsilon_{ab}h(r)e^{i\omega t}$ with the polarizations
$\varepsilon_{ab}$ analysed previously. 
Then, 
with respect to the action measure 
\beq
||h||^2 \sim \int {\rm d}r \sqrt{g}e^{-2\Phi}g^{ab}g^{cd}(  g^{rr}(\partial_r h_{ac}^{*}(r)
\partial_r h_{bd}(r) + \omega^2
 g^{tt}h_{ac}^{*}(r)h_{bd}(r)) \ ,
\eeq
$h_1(r)$ is normalizable while $h_2(r)$ is non-normalizable, 

Denote by $e^{i\omega r}$ an incoming wave and by $e^{-i\omega r}$
an outgoing wave.
From the expansion near infinity (\ref{exp1}) we see that $h_1(r)$ consists of both
incoming and outgoing
waves while $h_2(r)$ is an incoming wave. 
From the expansion near zero (\ref{exp2}) we see that both $h_1(r)$ 
and $h_2(r)$ are of the form of an incoming wave. 

In order to use the standard method of computing the absorption of the gravitons by the branes
we should construct a unique linear combination of the two solutions $h_1(r)$ and $h_2(r)$
by supplying appropriate boundary conditions.
For Dp branes with $p < 5$ (see the next section),
 this is achieved by requiring that the wave near $r=0$ will
be purely ingoing. However, as we can see from (\ref{exp2}), in our case  such a boundary condition
does not fix uniquely the linear combination of the two solutions.
We will, therefor, follow a different strategy. We will analyse the two solutions separately.
We will show that the first solution $h_1(r)$ does not lead to an absorption and is reflected
back from the brane completely, independently of the decoupling limit.
In contrast, we will see that second solution $h_2(r)$ leads to an absorption, and 
is absorbed completely by the brane, again independently of the decoupling limit.

Consider first the normalizable solution $h_1(r)$. We can read the scattering cross section 
from (\ref{exp1}) or  
alternatively we can compute the flux by
\beq
{\cal F} \sim \frac{1}{2i}\int  \sqrt{g}e^{-2\Phi}g^{ab}g^{cd}
g^{rr}(h_{ac}^*(r)\partial_r h_{bd}(r) -
\partial_r h_{ac}^*(h) h_{bd}(r))
\label{flux}
\ .
\eeq
We denote the incoming and outgoing fluxes at infinity by
${\cal F}_{\infty}^{in}$ and ${\cal F}_{\infty}^{out}$ respectively, and by
${\cal F}_{0}^{in}$ the incoming flux at zero.
We see that if we only consider the normalizable solution $h_1(r)$ we 
have
\beq
\frac{{\cal F}_{\infty}^{out}}{{\cal F}_{\infty}^{in}} = 1,~~~~~ {\cal F}_{0}^{in} = 0 \ .
\eeq
That means that there is no absorption of gravitons by the D6 branes and all the gravitons are
reflected back.
This is the familiar Rutherford scattering in a Coulomb potential. 
Indeed in the analysis of scattering in a Coulomb potential, only the normalizable
solution is relevant. 
The argument invoked in discarding the non-normalizable solution is that it requires
a $\delta$-function source at the origin, and such source does not exist.
Thus, there is no absorption by the point-like charged source at the
origin and all the waves are scattered back. 

This is not the physics of our system of D6 branes. At least before taking the decoupling limit
we expect gravitons to be absorbed by the branes and excite the brane modes such as the gauge
fields.
This implies that we have to consider the non-normalizable solution as well.
In the next section we will see that this is also required for other Dp brane when $p \neq 3$.
The reason why we do have to consider in all these cases the non-normalizable solution is
the fact that we have a singularity at $r=0$, due to the the Dp branes source.
Let us discuss this issue in some more detail.

The singularity of the D6 branes supergravity background
at $r=0$ is time-like. Having such a singularity of the
classical geometry which we can reach at a finite proper
time, there is the natural question whether
it is resolved quantum mechanically. One criterion \cite{HM} is the existence of a 
self-adjoint Laplacian. This can still be the case even if the metric
is geodesically incomplete. Indeed the requirement is the existence of 
a non-normalizable solution of the wave equation. This criterion is satisfied by our geometry.
Physically we expect that the singularity is resolved by the D6 branes source.
The details of the resolution may differ in the presence or absence of a $B$ field, since
it is related 
to the branes degrees of freedom which are missed by the classical supergravity background.
The field theory on the branes differ with a non-vanishing $B$ field compared to $B=0$. 
The non-normalizable solution is the  probe we have on the physics at $r=0$.

Consider now the non-normalizable solution $h_2(r)$.
We denote the incoming flux at infinity by
${\cal G}_{\infty}^{in}$ and by
${\cal G}_{0}^{in}$ the incoming flux at zero.
Again, we can compute the ratio of fluxes by using the wave function $h_2(r)$
or by computing the fluxes using  (\ref{flux}). In both cases we get
\begin{eqnarray}
  \label{Gflux}
{\cal G}_{0}^{in}&\sim&\frac{8}{\omega R^{2} |\Gamma(-\frac{i\omega R}{2})|^{2}}
\left[1+\omega R\;\frac{\pi}{2}-\omega R \;\;Im[\psi(1-\frac{i R\omega}{2})]\right],\\
{\cal G}_{\infty}^{in}&\sim&2\omega e^{\pi\omega R/2},
\end{eqnarray}
and 
\beq
\frac{{\cal G}_{0}^{in}}{{\cal G}_{\infty}^{in}} = 
\frac{2}{\pi\omega R}\sinh{(\frac{\pi\omega R}{2})}e^{-\frac{\pi\omega R}{2}}\left[1+\omega R\;\frac{\pi}{2}-\omega R\;Im[\psi(1-\frac{i \omega R}{2})]\right] 
\equiv 1 \ ,
\label{finaleq}
\eeq
where $Im[\psi]$ denotes the imaginary part of $\psi$,
and  $|\Gamma[ix]|^{2}=\frac{\pi}{x\sinh{(\pi x)}}$ has been used.

We see from (\ref{finaleq}) that  the non-normalizable solution $h_2(r)$
is completely absorbed by the brane.
This result does not change in the decoupling limit
$\omega l_s \rightarrow 0$, $\alpha_i\rightarrow \frac{\pi}{2}$
where we keep  
$b_i=l_s^2 \cos^{-1} \alpha_i$ and  $g_s l_s^{p-3-2m}$  fixed.
Therefore the graviton absorption computation indicates that gravity does not decouple
from the D6 branes worldvolume theory in the presence of a non-vanishing $B$ field.

\begin{figure}[htb]
\begin{center}
\epsfxsize=4in\leavevmode\epsfbox{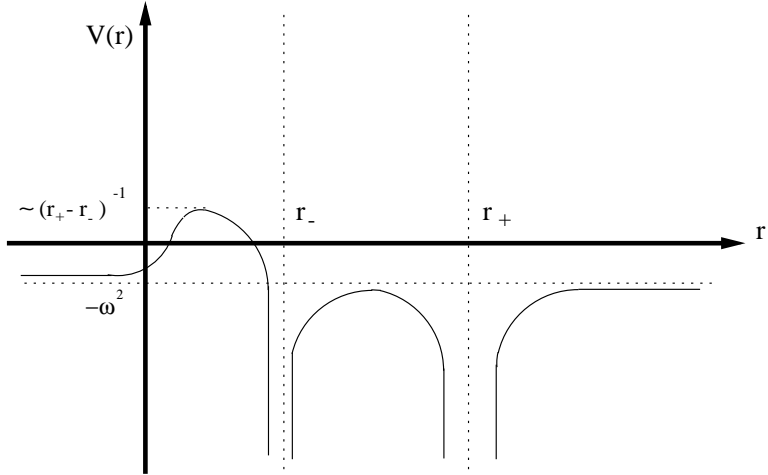}
\end{center}
\caption{The graviton scattering potential in the non-extremal D6 branes background.}
\label{nd6}
\end{figure}

We should note however that since
our computation has been done in the framework of supergravity there is still the
issue of the resolution of the singularity at $r=0$. 
One way to attempt at resolving  this singularity is to use the non extremal D6 branes 
solution \cite{HS} 
\begin{eqnarray}
ds^2&=&-(1-{r_{+}\over r})(1-{r_{-}\over r})^{-1/2}dt^2+
(1-{r_{+}\over r})^{-1}(1-{r_{-}\over r})^{1/2}dr^2 \cr &&\cr
&&(1-{r_{-}\over r})^{1/2}dx_i^2+r^2(1-{r_{-}\over r})^{3/2}d\Omega_2^2\ , \cr
&&\cr
e^{-2\phi}&=&(1-{r_{-}\over r})^{-3/2} \ .
\end{eqnarray}
As before, we 
decompose the graviton $h(r) = g(r)c(r)$ with
\begin{equation}
c(r)=\frac{1}{(r(r-r_{+}))^{1/2}} \ .
\end{equation}
Then 
the equation for $g(r)$ takes the form (\ref{sheq}) with the potential
\begin{equation}
V(r)=-\omega^2\;\frac{r(r-r_{-})}{(r-r_{+})^2}
-\frac{(r_{-}-r_{+})^2}{(r-r_{-})^2(r-r_{+})^2} \ 
\end{equation}
depicted in figure (\ref{nd6}).

The physical region is $r \geq r_+$.
The solutions of near $r=r_{+}$ are  $g(r) \sim(r-r_{+})^{\pm i\beta}$, where  
$\beta =\sqrt{1+\omega^2 r_{+}(r_{+}-r_{-})}$. At infinity the solutions take the form 
$g(r) \sim e^{\pm i \omega r}$. 
Both solutions are normalizable, and in the extremal limit $r_{+}=r_{-}$
we do not approach the non-normalizable solution. 
Thus, the proper way of resolving the singularity requires a better understanding
of the inclusion of the D6 branes degrees of freedom.
Details of such resolution may
depend on the $B$ field but it seems unlikely
that they will modify our conclusion about the non-decoupling of gravity.
For comparison we will analyse in the next section other cases where decoupling exists but 
there is a singularity at $r=0$.

\section{Graviton Scattering on Dp Branes} 

In this section we study the 
scattering of gravitons on  Dp-branes ($p<5$) in the supergravity
 description. We analyse the scattering potentials, the absorption cross section 
and discuss the issue of normalizability of scattering waves.
The purpose of this discussion is to compare with the D6 branes other 
cases where the bulk gravity does decouple from the brane
modes while the supergravity solution is singular.

\subsection{Dp Branes}

The Ricci scalar ${\cal R}_p$ and the dilaton $\Phi$
of the Dp branes supergravity background are 
\beq
{\cal R}_p \sim (p-3)\frac{R_p^{2(7-p)}}{r^{2(8-p)}f_p^{5/2}(r)},~~~~
 e^{\Phi} \sim f_p^{\frac{3-p}{4}}(r) \ ,
\eeq 
where 
$f_{p}=1+\frac{R^{7-p}_{p}}{(7-p)r^{7-p}}$ is the harmonic function and  
$R_{p}^{7-p} \sim g_sNl_s^{7-p}$.
When $p > 3$ the Ricci scalar diverges at $r=0$, while  when $p < 3$ 
the dilaton blows up.

As before,
the transversal graviton modes $h(r)$ are related to the minimally couple scalar $\varphi(r)$ like
\beq
h(r)=f_{p}^{-1/2}(r)\varphi(r) \ .
\eeq
Consider the minimally coupled scalars.
Scattering  waves 
\beq
\varphi(t,r,\Omega),=\phi_{l,m_{1},...,m_{7-p}}(r)
e^{i\omega t}Y_{l,m_{1},...,m_{7-p}}(\Omega)
\eeq
can be calculated by solving the differential equation  
\begin{eqnarray}
   \left[r^{-8 + p}\partial_{r} r^{8 - p} \partial_{r}+\omega^2\,f_{p}(r)-\frac{l(l+(7-p))}{r^{2}}\right] \phi_{l,m_{1},...,m_{7-p}}(r)=0,
\end{eqnarray}
for the radial factor $\phi(r)_{l,m_{1},...,m_{7-p}}$. 
Let $\phi_{l,m_{1},...,m_{7-p}}(\rho/\omega)=\psi(\rho)/(\omega^{2}\rho^{(8-p)/2)})$ 
and $\rho=\omega r$ one finds the Schr\"{o}dinger-like equation
\beq
\label{eq:differentialequation}
\partial_{\rho}^{2}\psi(\rho)-V_{p}(\rho)\psi(\rho)=0 \ ,\\
\eeq
where
\beq
 V_{p}(\rho)=
-\left(1+\frac{(\omega R_{p})^{7-p}}
{(7-p)\rho^{7-p}}\right)+{\frac{(p-8 )(p-6 ) }{{4 \rho^2}}}+\frac{l(l+7-p)}{\rho^{2}} \ .
\eeq

\begin{figure}[htb]
\begin{center}
\epsfxsize=4in\leavevmode\epsfbox{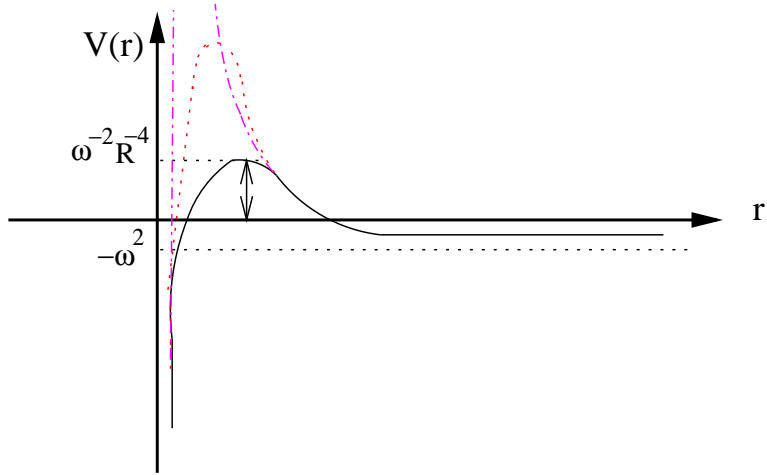}
\end{center}
\caption{The graviton scattering potential in Dp branes background. The value
of the height $\omega^{-2}R^{-4}$ corresponds to $p=3$.}
\label{dp}
\end{figure}

Let us discuss the potential $V_{p}(\rho)$, depicted in figure (\ref{dp}).
Consider for simplicity the s-wave. 
As $\rho$ approaches zero the potential diverges to minus infinity $V(\rho\rightarrow\infty)\sim-1/\rho^{7-p}$. As $\rho$ approaches infinity it converges to $-1$: $V(\rho\rightarrow\infty)=-1$. 
In between it reaches a maximum at
\begin{equation}
  \label{eq:maximum}
 \rho_{max}=\left(\frac{2(7-p)(\omega R_{p})^{7-p}}{(p-8)(p-6)}\right)^{\frac{1}{5-p}} \ .
 \end{equation}
In the decoupling limit $\omega R_{p}$ goes to
zero and the maximum value of the potential $V(\rho_{max})$ goes to infinity
\begin{equation}
  \label{eq:MaxPotential}
 V(\rho_{max})\sim \frac{1}{(\omega R_p)^{\frac{2(7-p)}{5-p}}},\quad \omega R_{p}\rightarrow0, 
\end{equation}
creating an infinitely high barrier.
In fact from $r=\infty$ to the radius $r_{max}$ the potential 
looks like the one of a free 
wave in flat space time, written in polar coordinates with $8-p$ angles. In the limit the potential barrier makes it impossible for scattering waves to reach $r=0$. The bulk modes decouple because of the potential around $r=r_{max}$. 
In comparison to the D6 branes we see that
there the barrier part is missing in the potential.

The absorption cross section per unit world volume of Dp branes ($p<5$) goes to zero in the 
decoupling limit. It is calculated in terms of fluxes of scattering solutions of (\ref{eq:differentialequation}).
\begin{eqnarray}
  \label{eq:crosssection}
  \sigma_{p}(\omega,R_{p})&=&\frac{(2\pi)^{8-p}}{\omega^{8-p}\Omega_{8-p}}\frac{{\cal F}^{in}_{0}}{{\cal F}^{in}_{\infty}},\quad \Omega_{d}=\frac{2\pi^{(d+1)/2}}{\Gamma(\frac{d+1}{2})},\\
   {\cal F}_{S}&=&\int_{S}J^{a}d\sigma_{a},\\
  {\cal F}_{S_{r=r_{S}}}&=&\int_{S_{r=r_{S}}}\sqrt{g}e^{-2\Phi}g^{rr}i(\partial_{r}{\varphi}^{*}\varphi-
{\varphi}^{*}\partial_{r}\varphi)dx^{p+1}_{\|}d\Omega_{8-p} \ .
 \end{eqnarray}
Here $S_{r=r_{S}}$ is the surface of all events of space time with $r=r_{S}$, $d\Omega_{8-p}$ is the volume element of the $(8-p)$-sphere.
 
It is difficult to write a closed solution to equation (\ref{eq:differentialequation}).
One can solve for $\rho\ll1$ and for $\rho\gg(R_{p}\omega)^{\alpha_{p}}$ ($\alpha_{p}>1)$. Assuming that $(R_{p}\omega)^{p-5}\ll1$, 
the two regions overlap and the asymptotic solutions can be matched
\footnote{Explicit calculations of these types can be found in \cite{KLEB}.}.
We summarize the results of the asymptotic solutions to the scattering equation 
in the background of Dp branes and the absorption
cross section  in the following table. 
\vskip 0.5cm
\begin{tabular}{|l||l|l|l|l|l|}\hline
p&$\rho\ll1$&$z$&$\rho\gg(R_{p}\omega)^{5-p}$&$\sigma_{p}(\omega,R_{p})$\\\hline\hline
1&$i(z)^{3/2}H^{1}_{3/2}(z)$&$(\omega R_{1})^{3}/\rho^{2}$&$48\sqrt{2/\pi}\rho^{-3}J_{3}(\rho)$&$\frac{2\pi^{4}}{3}\omega^{2}R_{1}^{9}$\\\hline

2&$i(z)^{5/3}H^{1}_{5/3}(z)$&$(2/3)(\omega R_{2})^{5/2}/\rho^{3/2}$&$c_{0}\rho^{-5/2}J_{5/2}(\rho)$&$c_{1}\omega^{2+1/3}R_{2}^{8+1/3}$\\\hline

3&$i(z)^{2}H^{1}_{2}(z)$&$(\omega R_{3})^{2}/\rho$&$\frac{32}{\pi}\rho^{-2}J_{2}(\rho)$&$\frac{\pi^{4}}{8}\omega^{3}R^{8}_{3}$\\\hline

4&$i(z)^{3}H^{1}_{3}(z)$&$2(\omega R_{4})^{3/2}/\sqrt{\rho}$&$24\sqrt{2/\pi}\rho^{-3/2}J_{3/2}(\rho)$&$\frac{2\pi^{3}}{3}\omega^{5}R_{4}^{9}$\\\hline
\end{tabular}
\vskip 0.5cm

 $H^{1}_{\nu}(z)=J_{\nu}(z)+iN_{\nu}(z)$, $J_{\nu}(z)$ and $N_{\nu}(z)$ denote the Hankel, Bessel and Neumann functions, respectively. (The constants $c_{0},c_{1}$ can be calculated to $c_{0}=\sqrt{\frac{\pi}{2}}\frac{5\;\;2^{11/3}}{\Gamma(1/3)\sqrt{3}}$, $c_{1}=\frac{\pi^{3}\Gamma(1/3)^{2}}{\sqrt[3]{3}\;2^{2}\;5}$).  
The linear combinations of the wave like solutions are uniquely determined by the physical boundary condition $F_{r=0}^{out}=0$. The absorption cross section in string units vanishes in the decoupling limit 
\begin{eqnarray}
  \label{eq:vanishingcrosssection}
  \sigma_{p}(\omega,R_{p})/l_{s}^{8-p}\rightarrow 0 \ .
\end{eqnarray}

The scattering solutions found above are not normalizable in the  norm induced from flux conservation
\begin{eqnarray}
  \label{eq:normdefinition}
  {\cal F}(S_{t=t_{S}})&=&\int_{S_{r=r_{S}}}\sqrt{g}e^{-2\phi}g^{tt}i(\partial_{t}\varphi^{*}\varphi
-\varphi^{*}\partial_{t}\varphi)dx^{p}_{\|}\hat{dx_{0}}drd\Omega_{8-p} \ .
\end{eqnarray}
Their current density blows up at $r=0$. Using the identities
\begin{eqnarray}
  \label{eq:asymptoticdensity}
  \phi(r)&\stackrel{r\rightarrow 0}{\sim}&z^{\nu}H_{\nu}^{1}(z),\\
  H_{\nu}^{1}(z\rightarrow\infty)&=&\sqrt{\frac{2}{\pi z}}e^{i(z-\pi\nu/2-\pi/4)}(1+O(\frac{1}{z})),\\
  \nu&=&\frac{7-p}{5-p},\quad z=\frac{5-p}{2}\frac{(\omega R_{p})^{(7-p)/2}}{\rho^{(5-p)/2}},\\
  \sqrt{g}e^{-2\Phi}g^{tt}(r)&=&f_{p}(r)r^{8-p}\Omega_{8-p}\stackrel{r\rightarrow 0}{\longrightarrow} r\frac{R^{7-p}_{p}}{(7-p)}\Omega_{8-p} \ ,
\end{eqnarray}
the flux density of ${\cal F}_{t}(r\rightarrow0)$ diverges as
\begin{eqnarray}
  \label{eq:blow up of density}
  \sqrt{g}e^{-2\Phi}g^{tt}i(\partial_{t}\varphi^{*}\varphi-\varphi^{*}
\partial_{t}\varphi)\stackrel{r\rightarrow 0}{\sim}r^{(5-p)/2}r r^{p-7}=\frac{1}{r^{(7-p)/2}} \ .
\end{eqnarray}
So it is not integrable for $p<5$. This is related to the fact that generalized eigen-functions 
are at most $\delta$-normalizable.

Let us sketch why a Gaussian wave packed at zero behaves nicely.
\begin{eqnarray}
  \label{eq:gaussianwave}
  \phi(r)\sim r^{\alpha}\int d\omega\frac{1}{\pi}e^{-(\omega-\omega_{0})^{2}}
 e^{i\omega(t+\frac{R_{p}^{(7-p)/2}}{r^{5-p}})}\sim r^{\tilde{\alpha}}
 e^{i\omega_{0}(t+\frac{R_{p}^{(7-p)/2}}{r^{p-5}})} e^{-(t+\frac{R_{p}^{(7-p)/2}}{r^{p-5}})^2}.
\end{eqnarray}
In this way at $r\rightarrow 0$ superposition produces a damping factor due 
to the negative power $\frac{R_{p}^{(7-p)/2}}{r^{p-5}}$ in the exponent. 
Thus, although the physical boundary conditions forced 
to construct non normalizable wave functions, appropriate superpositions resolve 
divergences. In fact this behaviour is due to the horizons being null.
Note in comparison that D6 wave solutions cannot be superposed to give a damping 
factor as above - it has a timelike horizon.

To summarize, although for the Dp branes with $p \neq 3$ the background
is singular  at $r=0$, when $p <5$ 
they differ from the D6 branes in several aspects. The scattering potential for the gravitons
in the Dp branes supergravity background has a barrier that goes to infinity
in the decoupling limit. The scattering potential in the D6 branes cases does not have such behaviour.
In both cases the non-normalizable solution leads to absorption. However, the absorption goes
to zero in the decoupling limit for the Dp branes but is nonzero in the D6 branes cases.
Furthermore  the non-normalizable solution is 
 $\delta$-normalizable for Dp branes but not for D6 branes.

The computation done in this section can be repeated with a rank $2m$
$B$ field. As in the
D6 branes cases, the scattering potential is the same with 
$R_p^{7-p} \sim g_sl_s N \prod_{i~odd}^{2m-1}\cos^{-1} \alpha_i $ \cite{AOS}.

\subsection{NS5 Branes}

\begin{figure}[htb]
\begin{center}
\epsfxsize=4in\leavevmode\epsfbox{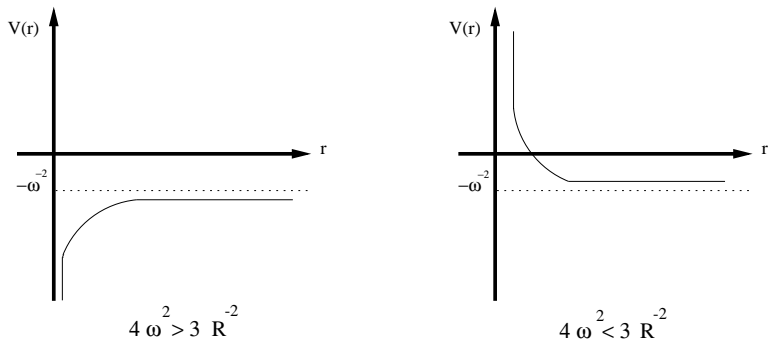}
\end{center}
\caption{ The scattering potential for gravitons in NS5 branes background.}
\label{NSfig}
\end{figure}

The case of Type II NS5 branes or D5 branes is different
from both Dp branes ($p< 5$) and D6 branes. In some sense this case is the border between these two cases. 
The scattering potential of the Schr\"{o}dinger-like  equation is
\begin{equation}
V(r)=-\omega^2+({3\over 4}-\omega^2 R^2)\frac{1}{r^2} \ ,
\end{equation}
where $R=Nl_s^2$. 
The sign 
of the second terms of
the potential changes at $\omega={\sqrt{3}\over {2R}}$ and therefore the 
shape of potential changes too as in figure \ref{NSfig}. The gravitons 
with
energy less than $\omega < {\sqrt{3}\over {2R}}$ see a barrier potential
and can not reach the branes. 

Setting  $\rho=\omega r$, the graviton scattering differential equation reads  
\begin{equation}
\frac{\partial^2 \psi}{\partial \rho^2}+\frac{3}{\rho}
\frac{\partial \psi}{\partial \rho}+\frac{R^2\omega^2}{\rho^2}\psi=0 \ .
\end{equation}
 The equation has solutions  $\rho^{\alpha}$, where 
\begin{equation}
\alpha=-1\pm \sqrt{1-(R\omega)^2} \ .
\end{equation}
For $R\omega >1$, where $\alpha$ becomes imaginary, there is a wave like
solution and. There is a non zero 
absorption cross section even in the decoupling limit \cite{MS} for the particles with
energy 
$\omega >\frac{m_s}{\sqrt{N}}$

The analysis of the graviton scattering for the NS5 branes in the presence of a  RR field is the same. 
As before, the only change is 
$R = N l_s^2 \prod_i \cos^{-1}\alpha_i$
\cite{MA,AOS}.

\section*{Acknowledgements}

We would like to thank A. Brandhuber,  J. Maldacena and G. Mandal 
for 
useful discussions.

\appendix{Type IIA Supergravity Equations}

The bosonic part of the low-energy effective action of the type IIA string theory is
\begin{eqnarray} 
S_{IIA}&=&\frac{1}{2\kappa^{2}}\int dx^{10} \sqrt{-G}\left(e^{-2\phi}\right.\left[R+4(\partial\phi)^{2}-\frac{1}{2\cdot 3!}H^{2}\right]-\frac{1}{2\cdot 2!}(F_{2})^{2}\nonumber\\
&&\quad\quad\quad\quad\quad\quad\quad\quad-\left.\frac{1}{2\cdot 4!}(\tilde F_{4})^{2}\right)-\frac{1}{4\kappa^{2}}\int \left.\left.B F_{4}F_{4}\right.\right. \ ,
\end{eqnarray}
where 
\begin{eqnarray}
F_{n}&=&dA_{n},\non
H&=&dB,\non
\tilde F_{4}&=&F_{4}-H\wedge A_{1} \ .
\end{eqnarray}
$B$ is the NS 2-form and $A_{n}$ denote the the RR $n$-form.
Varying the above action with respect to all the potentials we have:
\begin{eqnarray}
&&D^{k}F_{ki}=-\frac{1}{3!}F_{klmi}H^{klm}-\frac{1}{3!}H_{klm}H^{klm}A_{i}+\frac{1}{2!}H_{ilm}H^{lmn}A_{n},\\&&
D^{k}[F_{kijl}-4H_{[ijk}A_{l]}]=\frac{1}{3!\cdot4!}\varepsilon_{ijlmnopqrs}H^{mno}F^{pqrs},\\&&
D^{k}[e^{-2\phi}H_{kij}+H_{kij}A_{l}A^{l}-3A_{[i}H_{jk]l}A^{l}-F_{kijl}A^{l}]=\nonumber\\
&&\quad\quad\quad\quad\quad\quad\quad\quad\quad\quad\quad\quad\frac{1}{2\cdot4!\cdot 4!}\varepsilon_{kijlmnopqrs}F^{lmno}F^{pqrs},\\&&
D_{l}\partial^{l}\phi=-\frac{1}{4}R+\frac{1}{4\cdot2\cdot3!}H_{ijk}H^{ijk}+\partial_{k}\phi\partial^{k}\phi,\\&&
R_{ij}=-2D_{i}\partial_{j}\phi+\frac{3}{2\cdot3!}H_{(i}^{\;\;lm}H_{j)lm}\non
&&\quad\quad\quad\quad+e^{2\phi}\frac{1}{2\cdot2!}\left(2(F_{2})_{ik}(F_{2})_{j}^{\;\;k}-
\frac{1}{2}g_{ij}(F_{2})_{kl}(F_{2})^{kl}\right)\non
&&\quad\quad\quad\quad+e^{2\phi}\frac{1}
{2\cdot4!}
\left(4 (\tilde{F}_{4})_{(i}^{\;\;lmn}(\tilde{F}_{4})_{j)lmn}-\frac{1}{2}
g_{ij}(\tilde{F}_{4})_{klmn}(\tilde{F}_{4})_{klmn}\right).
\end{eqnarray}

The linearized field equations of the dilaton and the Ricci tensor take the form
\begin{eqnarray}
D_{l}\partial^{l}\phi-\Gamma(h)_{l}\partial^{l}\dot\phi
      -h^{lm}D_{l}\partial_{m}\dot\phi=\frac{1}{4}R_{ij}h^{ij}-\frac{1}{4}(D_{j}D_{i}h^{ij}-D^{2}h_{j}^{j})\nonumber\\
+\frac{2}{4\cdot2\cdot3!}H_{ijk}\dot{H}^{ijk}-\frac{3}{4\cdot2\cdot3!}h^{il}\dot H_{ijk}\dot{H}_{l}^{\;\;jk}
+2\partial_{k}\dot\phi\partial^{k}\phi-h^{kl}\partial_{k}\dot\phi\partial_{l}\dot\phi,
\end{eqnarray}
\begin{eqnarray}
&&D_{(i}D_{a}h^{a}_{\;\;j)}-\frac{1}{2}D^{2}h_{ji} -\frac{1}{2}D_{j}D_{i}h^{a}_{a}+(R_{g})_{m(i}h^{m}_{j)}+
(R_{g})_{jmni}h^{mn}= \nonumber\\
&&-2D_{i}\partial_{j}\phi +2\Gamma(h)_{i}\partial_{j}\dot\phi+\frac{2\cdot3}{2\cdot3!}\dot{H}_{(i}^{\;\;lm}H_{j)lm}-
\frac{2\cdot3}{2\cdot3!}h^{kl}\dot{H}_{(i|k}^{\;\;\;\;m}\dot{H}_{j)lm} \nonumber\\
&&+2\phi e^{2\dot\phi}\left(\frac{1}{2\cdot2!}[2(\dot F_{2})_{(i}^{\;\;l}(\dot F_{2})_{j)l}
-\frac{1}{2}g_{ij}(\dot F_{2})^{lm}(\dot F_{2})_{lm}]+ \nonumber\right.\\
&&\left.\quad\quad\quad\quad\frac{1}{2\cdot4!}[4(\dot{\tilde F}_{4})_{(i}^{\;\;lmn}(\dot{\tilde F}_{4})_{j)lmn}- 
\frac{1}{2}g_{ij}(\dot{\tilde F}_{4})^{lmno}(\dot{\tilde F}_{4})_{lmno}]\right)\nonumber\\
&&+e^{2\dot\phi}\left(\frac{1}{2\cdot2!}[2\cdot2(\dot F_{2})_{(i}^{\;\;l}(F_{2})_{j)l}-\frac{2}{2}g_{ij}(\dot F_{2})^{lm}(F_{2})_{lm}]\right.\nonumber\\
&&\left.\quad\quad\quad\quad+\frac{1}{2\cdot4!}[4\cdot2(\dot{\tilde F}_{4})_{(i}^{\;\;lmn}(F_{4}-\dot HA_{1}-H\dot A_{1})_{j)lmn}\nonumber\right.\\
&&\left.\quad\quad\quad\quad-\frac{2}{2}g_{ij}(\dot{\tilde F}_{4})^{lmno}(F_{4}-\dot H A_{1}-H\dot A_{1})_{lmno}]\right)\nonumber\\
&&-e^{2\dot\phi}\left(\frac{1}{2\cdot4!}[3\cdot4 h^{kl}(\dot{\tilde F}_{4})_{(i|k}^{\;\;\;\;mn}
(\dot{\tilde F}_{4})_{j)lmn}-\frac{4}{2}g_{ij}h^{kl}\dot{\tilde F}_{kmno}\dot{\tilde F}_{l}^{\;\;mno}] \nonumber\right.\\
&&\left.\quad\quad\quad\quad+\frac{1}{2\cdot2!}[2h^{kl}(\dot F_{2})_{(i|k}(\dot F_{2})_{j)l}-\frac{2}{2}g_{ij}h^{kl}(\dot F_{2})_{k}^{\;\;m}(\dot F_{2})_{lm}\nonumber\right.\\
&&\left.\quad\quad\quad\quad+\frac{1}{2}h_{ij}(\dot{\tilde F}_{4})^{lmno}(\dot{\tilde F}_{4})_{lmno}+\frac{1}{2}h_{ij}(\dot F_{2})^{lm}(\dot F_{2})_{lm}]\right) \ ,
\end{eqnarray}
where the dotted fields represent the background fields and
$\Gamma(h)$ stands for contractions of the $h_{ij}$ dependent part in the Christoffel 
symbols
\beq
(\Gamma(h))^{i}_{\;\;jk}=\frac{1}{2}g^{il}(D_{j}h_{lk}+D_{k}h_{lj}-D_{l}h_{jk})+O(h^{2}) \ .
\eeq
\appendix{Type IIB Supergravity equations}
The bosonic part of the low-energy effective action of the type IIB string theory is
\begin{eqnarray} 
S_{IIB}&=&\frac{1}{2\kappa^{2}}\int dx^{10} \sqrt{-G}\left(e^{-2\phi}\right.\left[R+4(\partial\phi)^{2}-\frac{1}{2\cdot 3!}H^{2}\right]-\frac{1}{2\cdot 3!}(\tilde{F}_{3})^{2}\nonumber\\
&&\quad\quad\quad\quad\quad\quad\quad\quad-\frac{1}{2}(F_{1})^{2}-\left.\frac{1}{4\cdot 5!}(\tilde{F}_{5})^{2}\right)-\frac{1}{4\kappa^{2}}\int \left.\left.A_{4}F_{3}H\right.\right.,
\end{eqnarray}
we impose  the self duality condition $F_{5}=*F_{5}$ at the level of the field equations, and 
\begin{eqnarray}
F_{n}&=&dA_{n},\non
H&=&dB,\non
\tilde F_{3}&=&F_{3}-H A_{0},\non
\tilde F_{5}&=&F_{5}-\frac{1}{2}(B\wedge F_{3}-A_{2}\wedge H).
\end{eqnarray}

Varying the above action with respect to all the potentials we have:
\begin{eqnarray}
&&\nabla_{l}\partial^{l}A_{0}=-\frac{1}{3!}H_{ijk}(F_{3}-A_{0}H)^{ijk},\\&&
\nabla^{k}(F_{3}-A_{0}H)_{kij}=\frac{1}{3!}(F_{5})_{ijklm}H^{klm},\\&&
\nabla_{k}[(e^{-2\phi}+(A_{0})^{2})H-A_{0}F_{3}]_{kij}=-\frac{1}{3!}(F_{5})_{ijklm}(F_{3})^{klm},\\&&
\nabla_{l}\partial^{l}\phi=-\frac{1}{4}R+\frac{1}{4\cdot2\cdot3!}H_{ijk}H^{ijk}+\partial_{k}\phi\partial^{k}\phi,\\&&
R_{ij}=-2D_{i}\partial_{j}\phi+\frac{3}{2\cdot3!}H_{(i}^{\;\;lm}H_{j)lm}\nonumber\\
&&\quad\quad\quad\quad+e^{2\phi}\frac{1}{2}\left((F_{1})_{i}(F_{1})_{j}-\frac{1}{2}g_{ij}(F_{1})^{k}(F_{1})_{k}\right)\nonumber\\
&&\quad\quad\quad\quad+e^{2\phi}\frac{1}{4\cdot4!}(\tilde{F}_{5})_{(i}^{\;\;lmnp}(\tilde{F}_{5})_{j)lmnp}\nonumber\\
&&\quad\quad\quad\quad+e^{2\phi}\frac{1}{2\cdot3!}\left[3(\tilde{F}_{3})_{(i}^{\;\;lm}(\tilde{F}_{3})_{j)lm}-\frac{1}{2}g_{ij}(\tilde{F}_{3})^{jlm}(\tilde{F}_{3})_{jlm}\right],\\&&
(\tilde{F}_{5})_{ijklm}=\frac{1}{5!}\varepsilon_{ijklmnopqr}(\tilde{F}_{5})^{opqrs},\\&&
6\:\:\partial_{[i}(\tilde{F}_{5})_{jklmn]}=\frac{6!}{3!\cdot3!}(F_{3})_{[ijk}(H_{3})_{lmn]}.
\end{eqnarray}

The linearized field equations of the dilaton and the Ricci tensor take the form
\begin{eqnarray}
D_{l}\partial^{l}\phi-\Gamma(h)_{l}\partial^{l}\dot\phi
      -h^{lm}D_{l}\partial_{m}\dot\phi=\frac{1}{4}R_{ij}h^{ij}-\frac{1}{4}(D_{j}D_{i}h^{ij}-D^{2}h_{j}^{j})\nonumber\\
+\frac{2}{4\cdot2\cdot3!}H_{ijk}\dot{H}^{ijk}-\frac{3}{4\cdot2\cdot3!}h^{il}\dot H_{ijk}\dot{H}_{l}^{\;\;jk}
+2\partial_{k}\dot\phi\partial^{k}\phi-h^{kl}\partial_{k}\dot\phi\partial_{l}\dot\phi,
\end{eqnarray}
\begin{eqnarray}
&&D_{(i}D_{a}h^{a}_{\;\;j)}-\frac{1}{2}D^{2}h_{ji} -\frac{1}{2}D_{j}D_{i}h^{a}_{a}+(R_{g})_{m(i}h^{m}_{j)}+(R_{g})_{jmni}h^{mn}=\nonumber\\
&&-2D_{i}\partial_{j}\phi+2(\Gamma(h))_{i}\partial_{j}\dot\phi+\frac{2\cdot3}{2\cdot3!}\dot{H}_{(i}^{\;\;lm}H_{j)lm}-\frac{2\cdot3}{2\cdot3!}h^{kl}\dot{H}_{(i|k}^{\;\;\;\;m}\dot H_{j)lm}\nonumber\\
      &&+2\phi e^{2\dot\phi}\left(\frac{1}{4\cdot4!}(\dot{\tilde{F}}_{5})_{(i}^{\;\;lmnp}(\dot{\tilde{F}}_{5})_{j)lmnp}+\frac{1}{2\cdot3!}\left[3(\dot{\tilde{F}}_{3})_{(i}^{\;\;lm}(\dot{\tilde{F}}_{3})_{j)lm}-\frac{1}{2}g_{ij}(\dot{\tilde{F}}_{3})^{klm}(\dot{\tilde{F}}_{3})_{klm}\right]\right.)\nonumber\\&&\quad\quad\quad\quad\left.+\frac{1}{2}\left[(\dot{F}_{1})_{(i}(\dot{F}_{1})_{j)}-\frac{1}{2}g_{ij}(\dot{F}_{1})^{k}(\dot{F}_{1})_{k}\right]\right)\nonumber\\&&
       +e^{2\dot\phi}\left(\frac{2}{4\cdot4!}(\dot{\tilde{F}}_{5})_{(i}^{\;\;lmnp}(F_{5}-\frac{1}{2}(B\dot{F}_{3}+\dot{B} F_{3}-A_{2}\dot{F}_{3}-\dot{A}_{2}F_{3}))_{j)lmnp}\right.\nonumber\\
&&\quad\quad\quad\quad\left.+\frac{1}{2\cdot3!}\left[3\cdot2(\dot{\tilde{F}}_{3})_{(i}^{\;\;lm}(F_{3}-A_{0}\dot{H}-\dot{A}_{0} H)_{j)lm}\right.\right.\nonumber\\
&&\quad\quad\quad\quad\quad\quad\quad\quad\left.\left.-\frac{2}{2}g_{ij}(\dot{\tilde{F}}_{3})^{klm}(F_{3}-A_{0}\dot{H}-\dot{A}_{0} H)_{klm}\right]\right.\nonumber\\
&&\quad\quad\quad\quad\left.+\frac{1}{2}\left[2(\dot{F}_{1})_{(i}(F_{1})_{j)}-\frac{2}{2}g_{ij}(\dot{F}_{1})^{k}(F_{1})_{k}\right]\right)\nonumber\\&&
       -e^{2\dot\phi}\left(\frac{4}{4\cdot4!}h^{kl}(\dot{\tilde{F}}_{5})_{(i|k}^{\;\;\;\;mnp}(\dot{\tilde{F}}_{5})_{j)lmnp}+\frac{1}{2\cdot3!}\left[2\cdot3h^{kl}(\dot{\tilde{F}}_{3})_{(i|k}^{\;\;\;\;\;m}(\dot{\tilde{F}}_{3})_{j)lm}\right.\right.\nonumber\\&&\quad\quad\quad\quad\left.\left.-\frac{3}{2}g_{ij}h^{kl}(\dot{\tilde{F}}_{3})_{k}^{\;\;mn}(\dot{\tilde{F}}_{3})_{lmn}+\frac{1}{2}h_{ij}(\dot{\tilde{F}}_{3})^{lmn}(\dot{\tilde{F}}_{3})_{lmn}\right]\right.\nonumber\\&&\quad\quad\quad\quad\left.+\frac{1}{2}\left[-\frac{1}{2}g_{ij}h^{kl}(\dot{F}_{1})_{k}(\dot{F}_{1})_{l}+\frac{1}{2}h_{ij}(\dot{F}_{1})^{l}(\dot{F}_{1})_{l}\right]\right).
\end{eqnarray}

\end{document}